\documentclass{PHYEAUTH}
\usepackage{graphicx}
\usepackage{amsmath}
\usepackage{amssymb}

\begin{document}
\begin{frontmatter}
\title{Spin injection and accumulation in inhomogeneous semiconductors}
\author{Dan~Csontos\thanksref{thank1}}
and
\author{Sergio~E.~Ulloa}
\address{Department of Physics and Astronomy, and Nanoscale and Quantum Phenomena Institute, Ohio University, Athens, OH 45701}
\thanks[thank1]{
E-mail: csontos@phy.ohiou.edu}

\begin{abstract}
We present a study of spin transport in charge and spin inhomogeneous semiconductor systems.
In particular, we investigate the propagation of spin-polarized electrons through a boundary between two
semiconductor regions with different doping concentrations. We use a theoretical and numerical method,
presented in this paper, based on a self-consistent treatment of a two-component version of the Boltzmann
transport equation. We show that space-charge effects strongly influence the spin transport properties, in
particular giving rise to pronounced spin accumulation and spin density enhancement.
\end{abstract}

\end{frontmatter}


All-semiconductor spintronics has recently become feasible due to the
availability of diluted magnetic semiconductors for spin injection\cite{fiederling,ohno}
and the long spin relaxation lengths present in
semiconductors. In order to realize semiconductor spintronic applications,
issues of spin injection, transport, manipulation and detection need to be
studied and understood. In particular, transport studies need to answer
questions pertaining to propagation and scattering of spins across
interfaces, effects of applied fields and inhomogeneous doping variations
and the resulting built-in electric fields. Some very interesting studies in
this direction have been published recently by several
groups\cite{zuticPRL,yuPRB,yuPRBlong,pershinPRL}. However, most of the investigations
have been based on drift-diffusion approaches and in some cases without
taking into account space-charge effects which can be very significant in
semiconductor transport.

In this paper we present a theoretical formulation beyond drift-diffusion
that is capable of describing charge and spin transport through strongly
inhomogeneous semiconductor systems, as well as nonequilibrium effects. Our
approach is based on the semiclassical Boltzmann transport equation, two
spin-dependent electron distribution functions and a self-consistent
description which allows us to fully take into account space-charge effects.
In the following, we will present our model and describe a numerical method
for the solution of the resulting non-linear system of differential
equations. Subsequently, we will exemplify the versatility of our model as
well as the importance of space-charge effects by calculating the spin
transport properties of a spin and charge inhomogeneous system, in
particular studying the transport across a doping interface. We show that
spin accumulation and magnification of the spin density imbalance occurs
around the space-charge region and compare our results with the charge
homogeneous case.

Our theoretical model is based on a two-component version of the Boltzmann
transport equation, in the relaxation-time approximation, as follows:
\begin{subequations}
\label{BTE}
\begin{eqnarray}
-\frac{eE(x)}{m^{\ast }}\frac{\partial f_{\uparrow }(x,v)}{\partial v}+v%
\frac{\partial f_{\uparrow }}{\partial x} &=&-\frac{f_{\uparrow
}(x,v)-f_{\uparrow }^{0}(x,v)}{\tau _{m}} \\
&-&\frac{f_{\uparrow }(x,v)-f_{\downarrow }(x,v)}{\tau _{sf}^{\uparrow
\downarrow }},  \notag  \label{BTEup} \\
-\frac{eE(x)}{m^{\ast }}\frac{\partial f_{\downarrow }(x,v)}{\partial v}+v%
\frac{\partial f_{\downarrow }}{\partial x} &=&-\frac{f_{\downarrow
}(x,v)-f_{\downarrow }^{0}(x,v)}{\tau _{m}} \\
&-&\frac{f_{\downarrow }(x,v)-f_{\uparrow }(x,v)}{\tau _{sf}^{\downarrow
\uparrow }},  \notag  \label{BTEdown}
\end{eqnarray}%
where $E(x)$ is the inhomogeneous electric field, $f_{\uparrow (\downarrow
)} $ is the electron distribution for the spin up(down) electrons, and where
we have introduced two scattering times, $\tau _{m}$ and $\tau _{sf}$, for
the momentum relaxation and spin flip times, respectively. The electron
distributions $f_{\uparrow (\downarrow )}^{0}$ are local equilibrium
distribution functions to which electrons with spin up(down) relax with the
scattering time $\tau _{m}$. In our calculations we assume nondegenerate
statistics and assume a Maxwell-Boltzmann distribution normalized with the
local density of spin up(down) electrons as our local equilibrium
distribution according to
\end{subequations}
\begin{equation}
f_{\uparrow (\downarrow )}^{0}=n_{\uparrow (\downarrow )}\left[ \frac{%
m^{\ast }}{2\pi k_{B}T}\right] ^{1/2}\exp (-m^{\ast }v^{2}/k_{B}T)~,
\label{MB}
\end{equation}%
where $T$ is the lattice temperature and $k_{B}$ is the Boltzmann constant.
The inhomogeneous field, $E(x)$, is coupled to the spin densities via the
Poisson equation%
\begin{equation}
\frac{d^{2}\phi }{dx^{2}}=-\frac{dE}{dx}=-e\frac{N_{D}(x)-n_{\uparrow
}(x)-n_{\downarrow }(x)}{\epsilon \epsilon _{0}}~,  \label{poisson}
\end{equation}%
where $\phi (x)$ is the electrostatic potential profile, $\epsilon $ is the
dielectric constant and $N_{D}(x)$ is the donor profile, where we in the
following assume unipolar transport, no acceptors and complete ionization of
the donors. The spin up(down) electron densities in eqs. (\ref{MB},\ref%
{poisson}) are obtained from the distribution functions $f_{\uparrow
(\downarrow )}(x,v)$ via%
\begin{equation}
n_{\uparrow (\downarrow )}(x)=\int f_{\uparrow (\downarrow )}(x,v)dv~.
\label{density}
\end{equation}

Equations (\ref{BTE}-\ref{density}) are coupled through the spin densities,
the electric field, and the spin flip scattering term in the BTE equations,
and thus, they need to be solved self-consistently. We use a numerical
approach based on finite difference and relaxation methods, that we
originally developed for the study of nonequilibrium effects in charge
transport through ultrasmall, inhomogeneous semiconductor channels\cite%
{csontosJCE},\cite{csontosAPL2005}. As boundary conditions, we adopt the
following scheme: For the potential, the values at the system boundaries are
fixed to $\phi (x_{l})=V_{b}$ and $\phi (x_{r})=0$ ($l,r$ denote the left
and right boundary of the sample, respectively), corresponding to an
externally applied voltage $V_{b}$. The \textit{electron charge density} is
allowed to fluctuate around the system boundaries subject to the condition
of global charge neutrality, which is enforced between each successive
iteration in the self-consistent Poisson-Boltzmann loop. The \textit{spin
density} at the boundary is determined by the degree of boundary
polarization $P=(n_{\uparrow }-n_{\downarrow })/(n_{\uparrow }+n_{\downarrow
})$, for which the density at the boundary is defined according to $%
n_{\uparrow (\downarrow )}=n/2(1\pm P)$. For an unpolarized boundary, at
which $n_{\uparrow }=n_{\downarrow }$, care must be taken regarding to
sample and/or contact size to ensure that any inhomogeneous spin density
within the sample has decayed such that $P=0$ is valid at the unpolarized
boundary. In addition, the size of the contacts has to be large enough, such
that the electric field deep inside the contacts is constant and low. This
allows us to use the analytical, linear response solution to the BTEs (\ref%
{BTE})
\begin{equation}
f_{\uparrow (\downarrow )}(x_{l,r},v)=f_{\uparrow (\downarrow
)}^{0}(x_{l,r},v)\left[ 1-vE(x_{l,r})\tau _{m}/k_{B}T\right] ,
\label{linBTE}
\end{equation}%
as phase space boundary conditions at $x_{l,r}$, where we use the local
equilibrium distribution, $f_{\uparrow (\downarrow )}^{0}(x_{l,r},v)$ and
local electric field, $E(x_{l,r})$, obtained from the previous numerical
solution to the Poisson-Boltzmann iterative loop. At the velocity cut-off in
phase space, we choose $f_{\uparrow (\downarrow )}(x,v_{\max })=f_{\uparrow
(\downarrow )}(x,-v_{\max })=f_{\uparrow (\downarrow )}^{0}(x,v)$, which is
reasonable since, in the calculations, we assume $v_{\max }\geq 30k_{B}T$. A
more detailed description and discussion of our numerical method (described
for pure charge transport) can be found in Ref. \cite{csontosJCE}.

In the following we apply our model for the study of spin transport through
a \emph{charge} and \emph{spin inhomogeneous} semiconductor structure. For
this purpose, we use a 5 $\mu $m long GaAs sample, across which we apply a
bias voltage $V_{b}=-0.3$ V. We assume that electrons are spin-polarized
with $P=1$ for $x<-0.1$ $\mu $m (the sample is defined for $-2.5\leq x\leq
2.5$ $\mu $m) and that other parameters in the calculations are $T=300$ K, $%
\tau _{m}=0.1$ ps, $\tau _{sf}=1$ ns and $\epsilon =13.1$. Furthermore, we
study two different structures, one charge homogeneous, with $N_{D}=10^{21}$
m$^{-3}$, and one charge inhomogeneous with $N_{D}=10^{21}$ m$^{-3}$ for $%
x<0.1$ $\mu $m, and 10$^{22}$ m$^{-3}$ for $x>0.1$ $\mu $m.
\begin{figure}[t]
\par
\begin{center}
\leavevmode \includegraphics[width=0.9\linewidth]{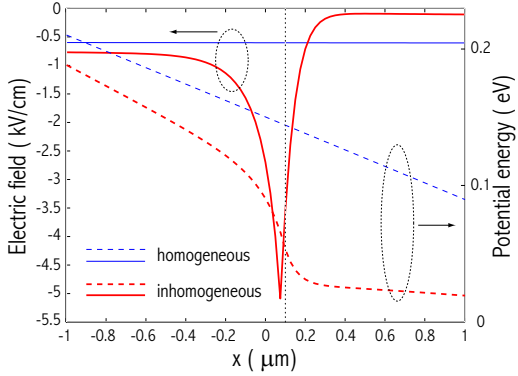}
\end{center}
\caption{(color online) Potential energy profiles (dashed lines) and
electric field distributions (solid lines) for a homogeneous sample with $%
N_{D}=10^{21}$ m$^{-3}$ (thin lines) and an inhomogeneous sample containing
a doping interface at $x=0.1$ $\protect\mu $m (vertical dotted line in the
figure) and doping concentrations of $N_{D}/N_{D}^{+}=10^{21}/10^{22}$ m$%
^{-3}$ (thick lines).}
\label{fig1}
\end{figure}

In Fig. \ref{fig1} we show the potential energy profiles (dashed lines) and
electric field distributions (solid lines) for the homogeneous (thin lines)
and inhomogeneous (thick lines) sample, respectively. A potential barrier is
formed at the interface between the two regions in the inhomogeneous sample,
as a consequence of electrons diffusing from the highly doped right region
to the lightly doped left region. Correspondingly, the electric field is
peaked around the interface and a space-charge region of $\approx 0.5$ $\mu $%
m is formed. Outside of this region, however, the electric field is constant
with $\left\vert E_{left}\right\vert >\left\vert E_{right}\right\vert $.
Naturally, for the homogeneous sample, the potential drops linearly over the
sample and the electric field distribution is constant, as illustrated by
the thin lines in Fig. \ref{fig1}.

\begin{figure}[t]
\par
\begin{center}
\leavevmode \includegraphics[width=0.9\linewidth]{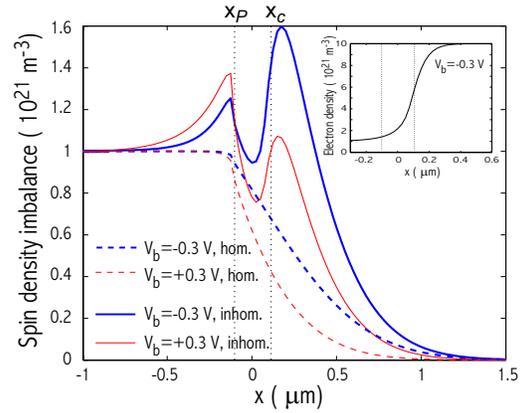}
\end{center}
\caption{(color online) Spin density imbalance, $\Delta n_{\uparrow
\downarrow }$, for the homogeneous (dashed lines) and inhomogeneous (solid
lines) case, calculated at $V_{b}=-0.3$ V (thick lines) and $V_{b}=0.3$ V
(thin lines). The interfaces between the unpolarized and polarized segments
occur at $x_{P}=-0.1$ $\protect\mu $m and $x_{c}=0.1$ $\protect\mu $m,
respectively, as indicated by the vertical dotted lines. Inset shows the
total charge density around the two interfaces, for $V_{b}=-0.3$ V. }
\label{fig2}
\end{figure}

In Fig. \ref{fig2} we show the calculated spin density imbalance, $\Delta
n_{\uparrow \downarrow }=n_{\uparrow }-n_{\downarrow }$, for the homogeneous
(dashed lines) and inhomogeneous structures (solid lines), calculated at $%
V_{b}=-0.3$ V (thick lines), and $V_{b}=0.3$ V (thin lines). We identify two
main features: First, it is evident that the calculated results for the
homogeneous and inhomogeneous samples differ dramatically around the
space-charge region. Second, the results differ significantly for opposite
sign of the bias voltage. The latter observation can be explained for the
homogeneous sample in terms of the findings of Yu and Flatt\'{e}\cite%
{yuPRB,yuPRBlong}, where the authors consider spin transport through a
homogeneous semiconductor structure in the presence of an applied electric
field.

Within this theoretical formulation, $\Delta n_{\uparrow \downarrow
}$ for our homogeneous structure can be described by $\sim \exp [-x/L_{d(u)}]
$ for $x>-0.1$ $\mu $m, where the $d(u)$ labels apply to the
negative(positive) bias case, and where $L_{d(u)}$ is the electric-field
dependent spin-diffusion length
\begin{equation}
L_{d(u)}=\left\{ -(+)\frac{\left\vert eE\right\vert }{2k_{B}T}+\sqrt{\left(
\frac{eE}{2k_{B}T}\right) ^{2}+\frac{1}{\left[ L^{(s)}\right] ^{2}}}\right\}
^{-1}~,  \label{spindiff}
\end{equation}%
where $L_{s}=\sqrt{D\tau _{sf}}$ is the intrinsic spin-diffusion length in
the absence of an electric field, and where $D=k_{B}T\tau _{m}/m^{\ast }$ is
obtained from the Einstein relation. From eq. (\ref{spindiff}) it follows
that the spin-diffusion length is enhanced in the direction anti-parallel to
an applied electric field and suppressed in the direction parallel to the
field.\cite{yuPRB},\cite{yuPRBlong} Hence, the difference between the two
dashed curves corresponding to the decay of $\Delta n_{\uparrow \downarrow }$
in the homogeneous sample, calculated at two bias voltages with opposite
sign, can be explained by an exponential decay with a field-dependent
diffusion length given by eq. (\ref{spindiff}).

The situation for the inhomogeneous sample is, however, very different. The
spin density imbalance, $\Delta n_{\uparrow \downarrow }$, has a
non-monotonic spatial dependence : \textit{i)} it increases before the
interface at $x<-0.1$ $\mu $m where the polarization is turned off, \textit{%
ii)} past the interface, for $x\geq -0.1$ $\mu $m, $\Delta n_{\uparrow
\downarrow }$ decreases, \textit{iii)} a second steep increase occurs around
the interface at $x=0.1$ $\mu $m between the $N_{D}=10^{21}$ m$^{-3}$ and $%
N_{D}=10^{22}$ m$^{-3}$ regions of the sample, where a peak is formed,
followed by a monotonic decrease toward 0 at the far-right side of the sample.

The origin of these features can be understood as follows: We can rewrite
the spin density imbalance according to
\begin{equation}
\Delta n_{\uparrow \downarrow }=n_{\uparrow }-n_{\downarrow
}=n-2n_{\downarrow }.  \label{spinimb}
\end{equation}%
Far to the left of the sample, where $P=1$, $n_{\downarrow }=0$.
Furthermore, the electric field and the charge density are constant and
correspondingly, $\Delta n_{\uparrow \downarrow }=n$. For a homogeneous
sample, only the second term in eq. (\ref{spinimb}) has a spatial variation
with a typical exponential decay as discussed above. However, \emph{in a
charge inhomogeneous structure, both the spin and charge densities have a
strong spatial dependence} and hence, both terms in eq. (\ref{spinimb})
affect the overall spatial dependence of $\Delta n_{\uparrow \downarrow }$.
The increase of $\Delta n_{\uparrow \downarrow }$ for $x<-0.1$ $\mu $m,
where $n_{\downarrow }\approx 0$, is due to a pure charge pile-up of the
total charge, $n$, which is increasing from left-to-right due to the
diffusion of electrons from the high-doping region to the low-doping region
to the left (see inset in Fig. \ref{fig2}). For $x>-0.1$ $\mu $m, spin
relaxation gives rise to an increase of $n_{\downarrow }$, the second term
in eq. (\ref{spinimb}), and hence reduces $\Delta n_{\uparrow \downarrow }$
as seen in the sudden drop in the spin density imbalance of Fig. \ref{fig2}.
However, around the interface at $x=0.1$ $\mu $m, $\Delta n_{\uparrow
\downarrow }$ increases again, and a sharp peak emerges.

This peak can be explained in terms of the spatial dependence of the total
charge $n$. Close to the doping interface at $x=0.1$ $\mu $m, there is a
sharp rise in the total charge density, as shown in the inset of Fig. \ref%
{fig2}, which occurs in order to accomodate the difference in doping
concentrations between the two regions. In this region, the increase of the
total charge $n$ is much faster than the increase in $n_{\downarrow }$ and
hence, $\Delta n_{\uparrow \downarrow }$ rises sharply, as given by eq. (\ref%
{spindiff}) and as seen in Fig. \ref{fig2}. Beyond the interface for $x>0.1$
$\mu $m, however, the total charge saturates at $n\approx 10^{22}$ m$^{-3}$
(see inset) and therefore, the gradual decrease of the $\Delta n_{\uparrow
\downarrow }$ peak for $x>0.1$ $\mu $m is solely due to the increase in $
n_{\downarrow }$ due to the spin-flip term in eq. (\ref{BTE}). We note that
spin accumulation at a doping interface has been recently reported by
Pershin and Privman \cite{pershinPRL}.

From the above discussion and results it is evident that space-charge
effects strongly influence the properties of semiconductor spin transport.
In particular, we conclude that spin transport characteristics depend on
several length scales, not only the electric-field dependent spin-diffusion
lengths defined in eq. (\ref{spindiff}), but also the charge screening
length, and the momentum relaxation length. Therefore, a self-consistent
treatment such as ours is needed for an accurate description of the
space-charge effects in semiconductor spintronics. We also note that the
importance of band-bending effects on spin injection in the nonlinear regime
of transport have been demonstrated in recent experiments \cite%
{schmidtPRL2004}.

\bigskip

This work has been supported by the Indiana 21st Century Research and
Technology Fund.

\end{document}